\documentstyle[11pt,aaspp4,psfig]{article}

\received{}
\received{}
\accepted{}
\cpright{ASP}{1998}

\slugcomment{submitted to \pasp}

\lefthead{Okamura et al.}
\righthead{Bulge and Disk Parameters from Growth Curve of Galaxies}

\begin{document}

\title{Retrieving Bulge and Disk Parameters and Asymptotic 
Magnitudes from the Growth Curves of Galaxies}

\author{Sadanori Okamura, Naoki Yasuda\altaffilmark{1}, Kazuhiro Shimasaku, 
and Masafumi Yagi\altaffilmark{1}}
\affil{Department of Astronomy and Research Center for the Early Universe,
School of Science, University of Tokyo, Bunkyo-ku, Tokyo, 113-0033 Japan;
okamura@astron.s.u-tokyo.ac.jp}

\and

\author{David H. Weinberg}
\affil{Department of Astronomy, Ohio State University, Columbus, OH,
43210, USA; dhw@astronomy.ohio-state.edu}

\altaffiltext{1}{Present address: National Astronomical Observatory,
Mitaka, Tokyo, 181-8588 Japan}

\begin{abstract} 
We show that the growth curves of galaxies can be used to determine
their bulge and disk parameters and bulge-to-total luminosity ratios,
in addition to their conventional asymptotic magnitudes,
provided that the point spread function is accurately known and 
signal-to-noise ratio is modest (S/N$\gtrsim30$). The growth curve is
a fundamental quantity that most future large galaxy imaging surveys will
measure. Bulge and disk parameters retrieved from the growth curve
will enable us to perform statistical studies of luminosity structure 
for a large number of galaxies.
\end{abstract}

\keywords{galaxies: photometry, growth curve, bulge-to-total ratio}

\section{Introduction}

The combination of a bulge and a disk is a conventional model to represent
the basic structure of galaxies.  The variety of surface brightness
profiles seen in galaxies is approximated reasonably well by assuming the
$r^{1/4}$-law, 
\begin{equation} 
\log \frac{I_B(r)}{I_{e,B}}=-3.33
	 \left[\left(\frac{r}{r_{e,B}}\right)^{1/4}-1\right],
\end{equation}
\noindent
for the bulge component and the exponential law,
\begin{equation} 
\log \frac{I_D(r)}{I_{e,D}} = -0.729
	 \left[\left(\frac{r}{r_{e,D}}\right)-1\right],
\end{equation} 
for the disk component. 
Here $I_e$ is the effective surface brightness and $r_e$ is the
major-axis effective radius. The subscripts B and D denote
bulge and disk, respectively. The effective surface brightness is
usually expressed in units of mag arcsec$^{-2}$ as 
\begin{equation} 
\mu_e = -2.5 \log I_e.  
\end{equation} 
The $r^{1/4}$-law was originally proposed to represent the profile of
elliptical galaxies. These ``laws'' do not have firm physical basis
(e.g., King \markcite{kin78} 1978; Freeman \markcite{fre70} 1970; Yoshii
\& Sommer-Larsen \markcite{yos89} 1989; Makino, Akiyama, \& Sugimoto
\markcite{mak90} 1990), and observed profiles of almost all galaxies
show, to some degree, departures from these laws (e.g., Kormendy
\markcite{kor77a} 1977a; Schombert \& Bothun \markcite{sch87b} 1987;
Shaw \& Gilmore \markcite{sg89} 1989). A number of authors (e.g.,
Kormendy \& Bruzual \markcite{kb78} 1978; Kent, Dame, \& Fazio
\markcite{kdf91} 1991;  Andredakis et al. \markcite{and95} 1995;
Courteau, de Jong, \& Broeils \markcite{cjb96} 1996) have given
evidence that many bulges can be fitted better by an exponential
radial profile than by an $r^{1/4}$-law.
Nevertheless, the empirical fitting
functions~(1) and~(2) 
are useful for characterizing and deriving global properties of galaxies.

The distributions and mutual relationships of the scaling parameters,
$\mu_{e,B}$, $r_{e,B}$, $\mu_{e,D}$, and $r_{e,D}$, are important clues
to the investigation of formation and evolution of galaxies.  Kormendy
\markcite{kor77a} (1977a) found a tight correlation between $\mu_e$ and
$r_e$ for elliptical galaxies.  A similar but slightly different
correlation holds for cD galaxies (Schombert \markcite{sch87} 1987).
Freeman \markcite{fre70} (1970) was the first to point out a very small
scatter of $\mu_{e,D}$, though he actually measured $\mu_D(0)$ instead
of $\mu_{e,D}$ in his study, using a sample of 36 spiral and S0
galaxies.  Disney \markcite{dis76} (1976) and Allen \& Shu
\markcite{all79} (1979) claimed that the near constancy of $\mu_{e,D}$
found by Freeman \markcite{fre70} (1970) is a result of selection
effects.  Indeed, later studies found many fainter disks (e.g., van der
Kruit \markcite{van87} 1987; Bothun et al. \markcite{bot87} 1987; Impey,
Bothun, \& Malin \markcite{imp88} 1988; Impey et al. \markcite{imp96}
1996; Dalcanton et al. \markcite{dal97} 1997), but brighter disks are
rare (McGaugh \markcite{mcg96} 1996).
The recent statistical study by  de Jong  \markcite{dej96b} (1996b)
has shown the bivariate distribution functions including
$\mu_{e,D}$, where we can see some kind of the Freeman's law,
but with a much larger scatter.

Our knowledge of the distributions and mutual relationships of the
scaling parameters is, however, still insufficient. This is mainly
because the samples analyzed so far are small and because the effects of
various selection biases are not understood very well (e.g., Kent
\markcite{ken85} 1985; Kodaira, Watanabe, \& Okamura \markcite{kod86}
1986; Simien \& de Vaucouleurs \markcite{sim86} 1986; McGaugh
\markcite{mcg96} 1996; de Jong \markcite{dej96b} 1996b; Dalcanton et al. 
\markcite{dal97} 1997).

Most previous studies were based on the so-called profile
decomposition technique, where the one-dimensional surface brightness
profile is extracted from two-dimensional surface photometry and the
bulge and the disk models are fit either iteratively or simultaneously
(e.g., Kormendy \markcite{kor77b} 1977b; Boroson \markcite{bor81} 1981; Kent
\markcite{ken85} 1985; Schombert \& Bothun \markcite{sch87b} 1987). 
A few authors proposed a two-dimensional image decomposition technique
(Shaw \& Gilmore \markcite{sg89} 1989;  de Jong \markcite{dej96a}
1996a). A method that depends on similar a priori fitting functions
was also applied to the study of dark halo problem
in spiral galaxies (e.g., Kent \markcite{ken86} 1986; Athanassoula, E., 
Bosma, A., \& Papaioannou, S. \markcite{abp87} 1987). 
Such decomposition techniques,
though they give relatively accurate parameters,
require an image of a
galaxy consisting of many pixels of high signal-to-noise ratio
(hereafter S/N).

Images of a large number of faint galaxies are available now (e.g.,
Impey et al. \markcite{imp96} 1996), and future CCD imaging surveys 
such as the Sloan Digital Sky Survey (hereafter SDSS; York
\markcite{yor98} 1998; 
Fukugita \markcite{fuk98} 1998) will increase
the number by more than an order of magnitude.  
If we wish to understand the distributions and correlations of
galaxy bulge and disk parameters and the implications of these
distributions for galaxy formation and evolution, it is essential to
derive parameters for these large samples and examine their statistical
behavior, even if the determination for any individual galaxy is
less accurate than could be obtained from a higher S/N image.

We show in this paper that the growth curve of galaxies is useful to
this purpose.  The growth curve is a plot of integrated magnitude as a
function of the diameter of the integrating (circular) aperture. The
growth curve is such a fundamental quantity that most future large
imaging surveys will measure it for a large number of galaxies.  The
growth curve has been conventionally used to derive the asymptotic
magnitude of galaxies by a method called growth curve fitting (e.g., de
Vaucouleurs et al. \markcite{dev76} 1976, RC2; de Vaucouleurs
\markcite{dev77} 1977; Kodaira et al. \markcite{kod90} 1990, PANBG; de
Vaucouleurs et al. \markcite{dev91} 1991, RC3).  We show that the growth
curve can be used, in addition to this conventional purpose, to infer
bulge and disk properties and bulge-to-total luminosity ratios of a
large number of faint galaxies, if the point spread function (hereafter
PSF) is accurately known and S/N is modest.

Since the original version of the present analysis was carried out in
the course of performance verification of the photometric pipeline
software of the SDSS, many parameters adopted in the present study are
tuned for the SDSS.  However, our conclusions are not subject to the
particular choice of the parameters.
We base our analysis on S/N, which is independent from the parameters
of the observation, and give a cross reference figure between
S/N and apparent magnitude that is valid for the SDSS.

We start in \S2 with a description of simulated galaxy images on which 
the present analysis is largely based. The method of growth curve fitting 
and construction of template growth curves are described in \S3.
Results from the simulated data are summarized in \S4. 
The effect of changing PSF is investigated in
\S5. Finally, a test of our method using images of real galaxies 
taken from PANBG is described in \S6.

\section{Simulated Galaxies}

We use two samples of simulated galaxies originally constructed to investigate
target selection criteria and photometric pipeline performance for the SDSS.
The bright sample
consists of 2110 galaxies with 12$\lesssim{r'}\lesssim18.5$ mag, while the
faint sample includes 2199 galaxies with 18.5$\lesssim{r'}\lesssim$23.3
mag. The bright sample and the faint sample are relevant to the
spectroscopic sample and the photometric sample of the SDSS, respectively.
The SDSS $r'$ band is centered at 6230\AA~ with a full width at half
maximum 1370\AA\ (see Fukugita \markcite{fuk1998} 1998)

Each simulated galaxy contains a bulge component and a disk component,
though for elliptical galaxies the disk component has zero luminosity.
The total $B$-band luminosities are drawn randomly from a Schechter
\markcite{sch76} (1976) luminosity function with parameters
$M_*=-19.68+5\log h$ and $\alpha=-1.07$.  Hubble types from E to Sc are
randomly assigned.  Bulge-to-disk luminosity ratios (in $r'$-band) are
assigned as a function of Hubble type based on Kent \markcite{ken85}
(1985).  K-corrected fluxes in the SDSS bandpasses are determined
separately for the bulge and disk components using the fiducial galaxy
spectral energy distributions of Coleman, Wu, \& Weedman
\markcite{col80} (1980).  Disk scale lengths $r_{e,D}$, are set using 
the Freeman \markcite{fre70} (1970) law and bulge scale lengths 
$r_{e,B}$ using the
fundamental plane relations as summarized by Maoz and Rix
\markcite{mao93} (1993), adjusted as a function of Hubble type based on
Kent \markcite{ken85} (1985).  A Gaussian scatter of 0.15 rms in
$\log r_e$ is added in order to produce a range of surface brightness at
each luminosity.  Disk axis ratios are assigned assuming random
inclinations, and bulge axis ratios are drawn randomly from the observed
distribution tabulated by Ryden \markcite{ryd92} (1992).  Galaxy
positions in a 3-dimensional model universe are taken from a large, cold
dark matter N-body simulation (by C. Park and J. R. Gott) in order to
produce an appropriately clustered galaxy population, but for purposes
of this paper each galaxy is analyzed in isolation and this clustering
is unimportant.
How to deal with  'blended images' is a separate problem, which we will
not discuss in the present paper.

For each galaxy in the input catalog, we create a simulated image in which
the bulge follows the $r^{1/4}$-law of equation
(1) and the disk follows the exponential law of equation (2). 
Figure 1 shows the distribution of $\log r_{e,B}$, $\log r_{e,D}$, and
$\eta=\log \left(r_{e,B}/r_{e,D}\right)$
of our 4309 sample galaxies. Most galaxies have 
$0 \farcs 1\lesssim{r_{e,B}},{r_{e,D}}\lesssim{10 \arcsec}$.

Our two-dimensional galaxy images have the characteristics expected for
the SDSS imaging survey; that is, we
simulate a CCD observation made at the F/5 focus of a 2.5-m telescope
in the $r'$ band with a resolution of 0.4 arcsec/pixel and an 
exposure time of 55 seconds. Sky brightness is assumed to be 21.2 mag
arcsec$^{-2}$.  The CCD plus filter response is also simulated
accordingly.  Statistical photon noise is added to the image.  
The image is convolved with a double-Gaussian seeing profile,
\begin{equation}
I(r)=0.9 \exp \left(-\frac{r^2}{2\sigma_1^2}\right)
   + 0.1 \exp \left(-\frac{r^2}{2\sigma_2^2}\right),
\end{equation}
where we assume $\sigma_1$=0.376 arcsec and $\sigma_2$=1.08 arcsec
to give a FWHM of $w_s=$0.944 arcsec. 

We define the signal-to-noise ratio, S/N, of a galaxy as
\begin{equation}
S/N = \frac{ \int [I(x,y)-I_{sky}]GdS}{\left[\int I(x,y)GdS\right]^{1/2}},
\end{equation}
where $I(x,y)$ represents the counts in a pixel
at $(x,y)$ in the image, $G$ is the gain factor, and integration is 
made within a circular aperture of diameter 10 arcsec centered on the image.
Figure 2 shows the S/N of 4309 galaxies as a function of apparent 
total magnitude $m_{cat}$ taken from the input catalog.

\section{Growth Curve Fitting and Template Growth Curves}

\subsection{Method of Growth Curve Fitting}

Growth curve fitting is a conventional method of galaxy photometry.
A growth curve is defined by
\begin{equation}
\Delta m(\xi)=m(\xi)-m_{asym},
\end{equation}
and
\begin{equation}
\xi=\log \left(\frac{r}{r_e}\right),
\end{equation}
where $m(\xi)$ is the magnitude of a galaxy integrated
in a {\it circular} aperture of radius $r$. The asymptotic
magnitude is defined by
\begin{equation}
m_{asym}=m(\infty).
\end{equation}
For the image of a galaxy with the center already identified,
we measure the magnitude of the galaxy integrated in a series
of circular apertures and obtain
\begin{equation}
 \{m_i(\xi_i); i=1, 2,....., N\}.
\end{equation}
The template growth curves are fit to the series of measurements
and the best-fit template is used to derive $m_{asym}$.
The {\it effective radius} $r_e$ is also obtained from the best-fit
template as
\begin{equation}
\Delta m(0)=0.75,\;\; \log r=\log r_e.
\end{equation}
Note that $r_e$ derived in this way is the effective radius
of the whole galaxy and not of an individual component.
Note also that  $r_e$ is the {\it circular-aperture
effective radius} and is different from the {\it major-axis effective radius}
that would be obtained by model fitting to the major-axis profile.

If a set of templates has a reasonable resolution in the 
scaling parameters, we can retrieve the bulge and disk parameters
from the best-fit template.  It is rather difficult to construct
such a set of growth curves based on real data. This is why we use
model galaxies instead of real data to construct the set of growth
curve templates, as described below.

\subsection{Template Growth Curves}

The conventional method of growth curve fitting relies on template
growth curves derived from real data. A template is usually made
for each morphological type index (de Vaucouleurs \markcite{dev77} 1977; RC2;
RC3), and in some cases for different inclinations as well (PANBG). 
It is important that the templates cover the full range 
of galaxy profiles.

Here we adopt a rather new approach of constructing the templates
using model galaxies that have a full range of realistic parameters.
A model galaxy for the template consists of a bulge and
a disk, which have the surface brightness distributions of
equations (1) and (2), respectively.
The templates used here are no longer the one-dimensional series of
morphological type index as in the previous studies.
We characterize the templates by two parameters. One is the
bulge-to-total luminosity ratio $B/T$, and the other is the
ratio of the effective radii $\eta=\log \left(r_{e,D}/r_{e,B}\right)$.

Another important factor is the effect of seeing. Seeing 
has different influence on galaxies with different apparent sizes.
In order to characterize the effect, we introduce another
parameter
\begin{equation}
\zeta=\log \left( r_{e,D}/w_s\right),
\end{equation}
where $w_s$ is the FWHM of the PSF (the seeing profile).
Assuming typical seeing of $w_s\sim1\arcsec$, we take
$-1<\log \zeta<1$, which is appropriate for galaxies concerned
(see Figure 1).

The ranges and the bin sizes of these parameters for 
the present set of templates are summarized in Table 1.
In total, 1089 (=$11\times9\times11$) templates are created.
Template growth curves are computed analytically for {\it face-on}
inclination using equations (1) and (2).
They are then convolved with the seeing profile of equation (4).
Convolution is performed in two
dimensions on the assumption of axisymmetry.
Figure 3 shows the 99 template growth curves before convolution with
the seeing profile.

\placetable{tbl-1}

\subsection{Practical Procedure}

In practice, for each galaxy, the sum of the squared residuals is computed 
for all the templates as
\begin{equation}
\delta^2=\sum_{i=3}^N w_i\left[ m(\log r_i)-m_{asym}-
         \Delta{m}\left(\log \frac{r_i}{r_e}(l,m,n)\right)\right]^2
        +\frac{1}{4}\left[ \log r_e - \log r_e(l,m,n)\right]^2,
\end{equation}
where $w_i$ is the weight and $(l,m,n)$ denotes the grid in the
$(B/T,\eta,\zeta)$ space. The second term of the right-hand side
is incorporated to avoid an apparently `good' fit by a template
with a very different $r_e$. In the present study, we adopt the
weight as
\begin{equation}
w_i = \log r_i.
\end{equation}
For a given set of $(l,m,n)$, the best-fit values of $m_{asym}$
and $\log r_e$ are computed by minimizing $\delta^2$.
The template that gives the smallest value of $\delta^2$ is chosen
as the best-fit template. We adopt the bulge and disk parameters of
the best-fit template as our estimates of their values.

This fitting procedure itself does not care about what kinds of 
models are used to create the template. Our choice of the 
$r^{1/4}$-law for the bulge component is a convention.
There is growing evidence that many bulges have steeper
surface brightness profiles than the $r^{1/4}$-law and that
they are better fitted by exponential profiles.
However, to examine the validity of these models is beyond
the scope of the present analysis. We would have similar
but slightly different parameters if we used exponential
models for the bulge component, as demonstrated by, 
for example, Andredakis \& Sanders \markcite{as94} (1994).
 
Table 2 shows the radii of our integrating apertures.
We follow the common practice of dropping
from the fitting procedure magnitudes that are integrated in a 
small aperture whose size is comparable to or smaller than
the seeing scale. In the present study, the innermost two magnitudes are 
not used in the growth curve fitting (see equation (12)).
Care should also be taken not to be affected badly
by occasional erratic behavior of the outermost data points due to
incorrect sky subtraction and/or excessive noise.

\placetable{tbl-2}

\section{Results}

\subsection{Asymptotic Magnitude}

In the measurement of $m_i(\xi_i)$ for each of the simulated galaxies, 
we reject data that either have S/N$<$1 or $m_i>m_{i-1}$. As a result,
279 galaxies in the faint sample yield fewer than four usable values
of $m_i(\xi_i)$,  
the minimum number required to derive the asymptotic magnitude with our 
scheme (equation 12).  Ten of these galaxies have three data values,
16 have two, six have one, and 247 have none.
Accordingly, asymptotic magnitudes are not determined for the 279 
galaxies.  These galaxies are very faint
and/or very small.

Figure 4(a) shows the magnitude difference
\begin{equation}
\delta{m}=m_{asym} - m_{cat},
\end{equation}
as a function of $m_{cat}$, where $m_{cat}$ is 
the total magnitude given in the input catalog, i.e., the correct answer.
Figure 4(b) shows the magnitude difference as a function of S/N.
It is seen in Figure 4(b) that $\delta{m}$(rms)$\lesssim0.1$ for 
S/N$\gtrsim30$ (log S/N$\sim$1.5). We hereafter take S/N=30 as the 
boundary between high-S/N galaxies and low-S/N galaxies.

In the case of some of the 
lowest-S/N galaxies (S/N$\lesssim$5), the best-fit template
gives an asymptotic magnitude that is very 
different from the true value ($>$2 mag). A typical example is shown in
Figure 5. If one wishes to pursue this method to such low-S/N galaxies
at the cost of relatively large error in the resulting asymptotic
magnitude, one could impose some constraint in order to circumvent 
the problem, for example, that
\begin{equation}
m_{asym} - m_{cat} > -2,
\end{equation}
when searching the best-fit template. 
However, $m_{cat}$ is, of course,
not available in the analysis of real data. We can replace $m_{cat}$
with an isophotal magnitude in this case.

\subsection{Bulge-to-Total Luminosity Ratio}

Figure 6(a) shows the correlation between the input and output
(measured) $B/T$  for high-S/N galaxies.
The input (catalog) $B/T$ values are sorted into 10 bins of
width $\Delta(B/T)=0.1$, and for each of the 10 bins the histogram
of the output $B/T$ values is shown in the separate panels
by the solid line.

There is a reasonably good correlation between the input and output
values. This is promising because it
suggests that we can estimate $B/T$ by growth curve fitting
for galaxies with high S/N. 
In the panels of high-$B/T$ bins ($B/T>0.5$), the histograms show
weak tails extending to low $B/T$ values. The tail can be seen even
in the highest bin ($B/T=1$). An inspection reveals that the 
tail mostly consists of galaxies that have small bulges with
$r_{e,B}\lesssim$1 pixel (0.4 arcsec). In the panels of low-$B/T$ 
bins ($B/T<0.2$),
on the other hand, tails are also seen to extend toward large $B/T$
values. This tail mostly consists of highly inclined galaxies
with $b/a\lesssim0.2$, which are very rare in reality.
In fact, only 264 (1.2\%) and 2043 (9.5\%) out of 21611 galaxies
in RC3 catalog have b/a$<0.1$ and  b/a$<0.2$, respectively.

It is reasonable that we cannot retrieve $B/T$ either for high-$B/T$
(bulge-dominated) galaxies with small $r_{e,B}$ or for low-$B/T$
(disk-dominated) galaxies with a highly inclined disk. In fact, if we
restrict our sample to galaxies with S/N$>30$ and $[ (B/T>0.5$ and
$r_{e,B}>1)$ or $(B/T<0.5$ and $(b/a)_D>0.2) ]$, we obtain a much
tighter correlation, as shown by the solid line in Figure 6(b).
However, it is impossible, of course, to restrict a sample of
observed galaxies according to their {\it intrinsic} parameters.
Figure 6(b) is just to show that our method works properly.

On the other hand, there is almost no
correlation for low-S/N galaxies (Figure 6(c)). This suggests that
little information on $B/T$, and hence on bulge and disk
parameters, can be retrieved for galaxies
with S/N$\lesssim$30.

The correlation shown in Figure 6(a) is promising for use in 
coarse morphological (structural) classification of galaxies.
In order to investigate quantitatively the use of the output $B/T$ 
for the morphological classification, we present the statistics
in Table 3, where the correlation matrix between output $B/T$ and
input $B/T$ is given for five classifications based on $B/T$
for four S/N classes. The mean and the standard deviation of
input $B/T$ ($\mu_i, \sigma_i$) and those of output $B/T$
($\mu_o, \sigma_o$) are also given. The former values,  
$\mu_i$ and $\sigma_i$  are determined by the input sample and 
do not have much significance.  We define the true fraction
by $N_{true}/N_m$, where $N_{true}$ is the number of galaxies
that are correctly measured in a given output $B/T$ bin and
$N_m$ is the number of all galaxies in the bin.  We compute
the true fraction in two ways; for $\Delta(B/T)=\pm0.1$ 
we count only the galaxies in the same output $B/T$ bin as the 
input $B/T$ bin while for $\Delta(B/T)=\pm0.3$ the 
difference of $\pm1$ bin is allowed for.
The latter case, $\Delta(B/T)=\pm0.3$, may correspond to
three coarse classifications as 'early', 'intermediate', and
'late', according to $B/T$.

It is remarkable that the true fraction for $\Delta(B/T)=\pm0.3$
is as high as $\gtrsim80$\% in S/N$>$30 and that it does not change 
very much as a function of S/N. However, the true fraction is significantly
lower in 30$>$S/N$>$10.
In the SDSS, S/N$\sim$30 corresponds to $r'\sim$19 mag (Figure 2). 
The number of galaxies observed in the SDSS
brighter than this magnitude will be of the order of $10^6$. It is very
important to obtain even coarse classification for such a large 
number of galaxies, and the method of growth curve fitting is a promising
tool for this purpose.

Figure 7 shows the mean of output $B/T$ ($\mu_o$) as a 
function of the mean of input $B/T$ ($\mu_i$) for the four 
S/N classes. The error bars, which are 
shown only for the highest and the lowest S/N classes for clarity, 
represent the standard deviations, $\sigma_o$ and $\sigma_i$. 
These values are taken from Table 3.
The three high S/N classes (S/N$>$30) show similar correlations between
$\mu_o$ and $\mu_i$. There is a systematic trend that $B/T$ is 
overestimated for $B/T{\rm (input)}\lesssim0.5$ but underestimated
for $B/T{\rm (input)}\gtrsim0.7$. This trend is due to the extended tails noted
above. The correlation for the lowest S/N class (S/N$<$30) 
is, however, considerably different, showing a larger degree of degradation 
in the output $B/T$. This marked change of behavior is the reason we
have chosen S/N=30 as the boundary between high-S/N and low-S/N galaxies.

\placetable{tbl-3}

\subsection{Bulge and Disk Parameters}

Figure 8 shows the correlations between the input and output values
for $r_{e,B}$ and $r_{e,D}$. For $r_{e,B}$, only the galaxies with
$B/T>$0.2 are plotted, while for $r_{e,D}$, only those with
$B/T<$0.8 are plotted.
The filled circles and small dots are for face-on galaxies
($(b/a)_B\geq0.8$ and $(b/a)_D\geq0.8$) and 
inclined galaxies ($(b/a)_B<0.8$ or $(b/a)_D<0.8$), respectively.
Left panels are for high-S/N galaxies and right panels 
for low-S/N galaxies.

Low-S/N galaxies show almost no correlation, as expected from the
$B/T$ statistics. However, high-S/N galaxies show reasonably
tight correlations. It is seen that $r_{e,B}\sim$1 pixel (0.4 arcsec;
log $r_{e,B}\sim -0.4$) is the limit below which we cannot retrieve
the bulge parameters, as expected from the analysis of $B/T$.
Statistics are shown in Table 4 in the similar format
to Table 3.

Face-on galaxies, which should have better match with the 
templates, show fairly good correlations. However, it is
interesting to see that most of the inclined 
galaxies also show correlations that are only slightly poorer
than those of face-on galaxies.
This indicates that growth curve fitting is not very sensitive
to galaxy inclination.

\section{Effect of Changing PSF}

It is natural that the retrieval of $B/T$ and the bulge/disk
parameters from the best-fit template is sensitive to the 
shape of the PSF. We investigate this effect
by applying an artificially changed PSF to the templates
while keeping the true PSF applied to the simulated galaxies.
Considering the practical application, here we restrict ourselves 
to a slight change of the PSF between the templates and the data, 
which would easily result from errors in the construction of the PSF. 
A much larger difference in the PSF shape would clearly jeopardize
the analysis, but it is reasonable to assume that with good
imaging data the PSF will be known to good, but not perfect, accuracy.

The PSF is usually constructed by summing up 
stellar images in a CCD frame by the shift and add method.
Such a PSF may well show a slight departure from the true
PSF due to image deterioration.
We created a large number of simulated (noisy) images of a 21 
mag star. We intentionally chose among them four images 
that have smaller FWHM than the average and reconstructed
the PSF using the four by a conventional shift-and-add method.
The resulting PSF is slightly but systematically different from 
the true (noiseless) PSF.
The reconstructed PSF has a FWHM of 0.919 arcsec while
the true PSF has FWHM of 0.944 arcsec.
The two PSFs and their profile difference are shown in Figure 9. 
The profile difference is less than 30\% in the core of the PSF but 
amounts to $\gtrsim$ 50\% in the envelope. However, their difference
in the FWHM is only 2.6\%.

The histogram of $B/T$ obtained with this reconstructed PSF is
shown for high-S/N galaxies in Figures 6(a)-6(c) by the broken line. 
We find only a little difference. Accordingly, $B/T$ statistics
of such galaxies are shown to be insensitive to change of 
the PSF at a few \% level in FWHM.
Correlations between the input and output values
for bulge and disk parameters also show only a slight change.

\section{A Test using Real Images of Bright Galaxies}

As a test on real data, we
apply the technique to images of nearby galaxies taken from PANBG.
There are 103 galaxies in PANBG whose $B/D$ ratios are measured
by Kodaira et al. \markcite{kod86} (1986), where $B/D$ is the bulge-to-disk
luminosity ratio and $B/T=(1-1/(1+B/D))$. The magnitudes of
these galaxies are $m_B$=9--12 mag. Their images, which
are digitized images of Schmidt plates, usually include more than
10$^4$ pixels above the detection threshold. The seeing FWHM 
of these data is 1--4 pixels, which is negligibly small compared with 
the size of the galaxies.

We measure their images in the same manner as the simulated images. 
Figure 10 shows the comparison of the $B/T$.
There is a reasonably good correlation between the two 
completely independent measurements. Most measurements agree with 
each other within $\Delta(B/T)=\pm0.1$. Figure 11 shows
our $B/T$ as a function of the morphological type index.
There is also a reasonably good correlation. 
The scatter between $B/T$ and type index is substantial,
but most previous investigations also show similar scatter
(e.g., Kodaira, Watanabe, \& Okamura \markcite{kod86} 1986;
Andredakis and Sanders \markcite{as94} 1994; de Jong
\markcite{dej96b} 1996b).

\section{Conclusion and Discussion}

We have shown that the growth curve of galaxies is useful for
determining bulge and disk parameters and bulge-to-total
luminosity ratios in addition to conventional asymptotic magnitudes.
Simulated images of 4309 galaxies with a wide
range of bulge and disk parameters are created.
Template growth curves are constructed from model galaxies
that cover the full range of realistic bulge and disk parameters.
The effect of seeing is also modeled in the growth curve templates.

We find that the bulge-to-total luminosity ratio and
bulge and disk parameters are reasonably well retrieved,
in addition to the conventional asymptotic magnitude (rms error
$\lesssim$0.1 mag), for galaxies with S/N$\gtrsim30$, if the error 
of PSF is $\lesssim$ a few percents in FWHM.
A coarse, three-bin $B/T$ classification can be made at a 
confidence level of more than 80 percent for galaxies with
S/N$\gtrsim$30.

Finally, it should be noted that the present analysis gives rather
optimistic estimates for the accuracy of this method for two reasons.
First, our simulated images and the template growth curves
are constructed from the same series of models, consisting
of an $r^{1/4}$-law bulge and an exponential disk.
Real galaxies often show slight departures from the model
as noted already. Such departures may well introduce some
error in the estimate of the parameters. However, the growth
curve is not sensitive to such slight departures because it is
an integrated profile, so template mismatch should not be a severe
problem at least in the case of regular galaxies.
Second, envelopes of bright stars sometimes overlap on 
faint galaxies in the actual survey data. They make sky
subtraction difficult and introduce systematic error,
which is difficult to quantify accurately.
Both of these potential difficulties will have to be
investigated in the context of individual large imaging data sets,
such as the SDSS.  However, the success of our tests here suggests
that growth curve fitting will be a valuable tool in the analysis
of such data sets, aiding the passage from pixel data to statistical
characterization and physical understanding of the galaxy population.

\acknowledgments

We thank M.Doi, Y.Komiyama, and W.Kawasaki for useful discussion
on many subjects related to this paper.
We thank Michael Strauss for assistance in calculating the K-corrected
fluxes for the artificial galaxy catalog and Changbom Park and 
Richard Gott for contributing the N-body simulation used to 
construct this catalog.
Special thanks go to the anonymous referee, who made many valuable
comments that led to significant improvements of
the manuscript.
This work was supported in part by Grants-in-Aid (07CE2002,
09640312) from the Ministry of Education, Science, Sports, and 
Culture of Japan.

\clearpage

\clearpage

\begin{deluxetable}{cccc}
\tablecaption{Parameters for the template growth curves. \label{tbl-1}}
\tablewidth{0pt}
\tablehead{
\colhead{Parameter} & \colhead{range} & \colhead{bin size} &
\colhead{number of templates}
}
\startdata
$B/T$                                   & \phs0.0 - 1.0 & 0.1 & 11 \nl
$\eta=\log\left(r_{e,B}/r_{e,D}\right)$ & $-$1.0 - 0.6 & 0.2 & \phn9 \nl
$\zeta=\log\left(r_{e,D}/w_s\right)$    & $-$1.0 - 1.0 & 0.2 & 11 \nl
\enddata
\end{deluxetable}

\clearpage

\begin{deluxetable}{ccccccccc}
\tablenum{2}
\tablecaption{Radii of the integrating aperture. \label{tbl-2}}
\tablewidth{0pt}
\tablehead{
\colhead{$i$} & \colhead{0} & \colhead{1} & \colhead{2} & \colhead{3} &
\colhead{4} & \colhead{5} & \colhead{6} & \colhead{7}
}
\startdata
$r_i$(pix)     &0.56  &1.69  &2.58  &4.41  &7.51  &11.58  &18.58  &28.55 \nl
$r_i$(\arcsec) &0.224 &0.676 &1.032 &1.764 &3.004 & 4.632 & 7.432 &11.42 \nl
\hline
$i$ & 8 & 9 & 10 & 11 &
12 & 13 & 14 &  \nl
\hline
 &45.50 &70.51 &110.5 &172.5 &269.5 &420.5 &657.5 & \nl
 &18.2  &28.2  & 44.2 & 69.0 &107.8 &168.2 &263.0 & \nl
\hline
\enddata
\end{deluxetable}

\clearpage

\begin{deluxetable}{crcrrrrrcc}
\tablenum{3}
\tablecolumns{10}
\tablewidth{0pt}
\tablecaption{Morphological classification based on $B/T$ ratio. \label{tbl-3}}
\tablehead{
\colhead{output} & \colhead{} & \colhead{} &
\multicolumn{5}{c}{input $B/T$} & &  \\
\cline{4-8}
\colhead{$B/T$} & \colhead{$N_m$} & \colhead{} & \colhead{0.0-0.2} &
\colhead{0.2-0.4} & \colhead{0.4-0.6} & \colhead{0.6-0.8} & 
\colhead{0.8-1.0} &  \multicolumn{2}{c}{True Fraction}  \\
}
\startdata
{(S/N$>$200)}&  &$\mu_i$   & 0.12 & 0.30 & 0.50 & 0.70 & 0.98  &
\multicolumn{2}{c}{$\Delta(B/T)$} \\
\cline{9-10}     
  &    &$\sigma_i$& 0.06 & 0.05 & 0.06 & 0.06 & 0.06  &
 $(\pm0.1)$ & $(\pm0.3)$ \\   
\cline{4-8}
 0.0-0.2 & 27& & 18&   9&   0&   0&   0&  67\%&  100\% \nl
 0.2-0.4 & 62& & 16&  30&  14&   2&   0&  48\%&   97\% \nl
 0.4-0.6 &132& & 11&  29&  58&  31&   3&  44\%&   89\% \nl
 0.6-0.8 &125& &  2&   6&  26&  71&  20&  57\%&   94\% \nl
 0.8-1.0 & 53& &  0&   1&   1&   5&  46&  87\%&   96\% \nl
\cline{4-8}
total& 399&$\mu_o$    & 0.33 & 0.45 & 0.58 & 0.69 & 0.87 &   &  \nl
     &    &$\sigma_o$ & 0.17 & 0.16 & 0.13 & 0.10 & 0.10 &   &  \nl
\tableline
{(200$>$S/N$>$100)}&  &$\mu_i$   & 0.10 & 0.30 & 0.50 & 0.71 & 0.98  &
\multicolumn{2}{c}{$\Delta(B/T)$} \\
\cline{9-10}
      &  &$\sigma_i$& 0.05 & 0.06 & 0.06 & 0.06 & 0.06  &
 $(\pm0.1)$ & $(\pm0.3)$ \\ 
\cline{4-8}
 0.0-0.2 & 71&  & 50&  16&   2&   1&   2&  70\%&   93\% \nl
 0.2-0.4 &175&  & 60&  64&  30&  18&   3&  37\%&   88\% \nl
 0.4-0.6 &271&  & 25&  62&  90&  75&  19&  33\%&   84\% \nl
 0.6-0.8 &171&  &  6&  16&  22&  98&  29&  57\%&   87\% \nl
 0.8-1.0 & 70&  &  1&   1&   1&  10&  57&  81\%&   96\% \nl
\cline{4-8}
total& 758&$\mu_o$    & 0.34 & 0.46 & 0.53 & 0.65 & 0.79 &   &  \nl
      &   &$\sigma_o$ & 0.17 & 0.16 & 0.12 & 0.14 & 0.19 &   &  \nl
\tableline
{(100$>$S/N$>$30)}&  &$\mu_i$   & 0.10 & 0.28 & 0.49 & 0.70 & 0.99  & 
\multicolumn{2}{c}{$\Delta(B/T)$} \\
\cline{9-10}
  &    &$\sigma_i$& 0.06 & 0.06 & 0.06 & 0.06 & 0.05  &
 $(\pm0.1)$ & $(\pm0.3)$ \\ 
\cline{4-8}
 0.0-0.2 &143&  &102&  23&   5&  10&   3&  71\%&   87\% \nl
 0.2-0.4 &265&  & 89& 100&  43&  31&   2&  38\%&   88\% \nl
 0.4-0.6 &351&  & 59&  78& 102&  93&  19&  29\%&   78\% \nl
 0.6-0.8 &256&  & 13&  14&  39& 118&  72&  46\%&   89\% \nl
 0.8-1.0 & 61&  &  0&   1&   2&   9&  49&  80\%&   95\% \nl
\cline{4-8}
total&1076&$\mu_o$    & 0.33 & 0.43 & 0.54 & 0.61 & 0.78 &   &  \nl
     &    &$\sigma_o$ & 0.18 & 0.15 & 0.15 & 0.16 & 0.16 &   &  \nl
\tableline
{(30$>$S/N$>$10)}&  &$\mu_i$    & 0.10 & 0.31 & 0.51 & 0.71 & 0.99 & 
\multicolumn{2}{c}{$\Delta(B/T)$} \\
\cline{9-10}
  &    &$\sigma_i$ & 0.06 & 0.05 & 0.07 & 0.05 & 0.04 & 
 $(\pm0.1)$ & $(\pm0.3)$ \\
\cline{4-8}
 0.0-0.2 & 42&  & 12&   6&   4&  16&   4&  29\%&   43\% \nl
 0.2-0.4 & 55&  &  5&   6&   3&  26&  15&  11\%&   25\% \nl
 0.4-0.6 & 90&  &  8&   6&   6&  40&  30&   7\%&   58\% \nl
 0.6-0.8 & 72&  &  8&   1&   4&  28&  31&  39\%&   88\% \nl
 0.8-1.0 & 11&  &  0&   0&   0&   2&   9&  82\%&  100\% \nl
\cline{4-8}
total& 270&$\mu_o$    & 0.40 & 0.39 & 0.47 & 0.51 & 0.60 &   &  \nl
     &    &$\sigma_o$ & 0.27 & 0.20 & 0.21 & 0.21 & 0.20 &   &  \nl
\enddata
\end{deluxetable}

\clearpage

\baselineskip=10pt
\begin{deluxetable}{crcrrrrr}
\tablenum{4a}
\tablecolumns{8}
\tablewidth{0pt}
\tablecaption{Correlation statistics for the bulge scale length. \label{tbl-4a}}
\tablehead{
\colhead{output} & \colhead{} & \colhead{} &
\multicolumn{5}{c}{input log $r_{e,B}$}  \\
\cline{4-8}
\colhead{log $r_{e,B}$} & \colhead{$N_m$} & \colhead{} & 
\colhead{-3.0 - -1.0} & \colhead{-1.0 - -0.5} & 
\colhead{-0.5 -  0.0} & \colhead{ 0.0 -  0.5} & 
\colhead{ 0.5 -  1.6} \\
}
\startdata
{(S/N$>$200)}&  &$\mu_i$   & -1.49& -0.67& -0.22& 0.27& 0.80 \nl
           &    &$\sigma_i$&  0.37&  0.12&  0.13& 0.13& 0.23 \nl
\cline{4-8}
-3.0 - -1.0&  0& &   0&   0&   0&   0&   0 \nl
-1.0 - -0.5&  1& &   0&   1&   0&   0&   0 \nl
-0.5 -  0.0& 71& &   1&   4&  43&  22&   1 \nl
 0.0 -  0.5&154& &   2&   4&  11& 123&  14 \nl
 0.5 -  1.6&173& &   4&   0&   6&  15& 148 \nl
\cline{4-8}
total& 399&$\mu_o$    & 0.48&-0.02& 0.01& 0.29& 0.79  \nl
     &    &$\sigma_o$ & 0.41& 0.36& 0.29& 0.18& 0.28  \nl
\tableline
{(200$>$S/N$>$100)}&   &$\mu_i$    &-1.37&-0.67&-0.21& 0.26& 0.73  \nl
                  &    &$\sigma_i$ & 0.27& 0.13& 0.14& 0.14& 0.18  \nl
\cline{4-8}
-3.0 - -1.0&  4& &   0&   4&   0&   0&   0 \nl
-1.0 - -0.5& 16& &   1&   6&   9&   0&   0 \nl
-0.5 -  0.0&209& &   1&  26& 120&  62&   0 \nl
 0.0 -  0.5&290& &   2&  17&  47& 195&  29 \nl
 0.5 -  1.6&239& &   9&   9&  30&  65& 126 \nl
\cline{4-8}
total& 758&$\mu_o$    & 0.74&-0.01& 0.05& 0.29& 0.69 \nl
     &    &$\sigma_o$ & 0.67& 0.50& 0.36& 0.26& 0.23 \nl
\tableline
{(100$>$S/N$>$30)}&    &$\mu_i$    &-1.61&-0.69&-0.23& 0.25& 0.71 \nl
                  &    &$\sigma_i$ & 0.50& 0.14& 0.14& 0.14& 0.16 \nl
\cline{4-8}
-3.0 - -1.0& 10& &   1&   8&   1&   0&   0 \nl
-1.0 - -0.5& 19& &   1&   7&  10&   1&   0 \nl
-0.5 -  0.0&266& &   8&  45& 130&  82&   1 \nl
 0.0 -  0.5&380& &  10&  22&  93& 215&  40 \nl
 0.5 -  1.6&401& &  20&  18&  54& 138& 171 \nl
\cline{4-8}
total&1076&$\mu_o$    & 0.43&-0.03& 0.15& 0.35& 0.71 \nl
     &    &$\sigma_o$ & 0.54& 0.52& 0.38& 0.29& 0.25 \nl
\tableline
{(30$>$S/N$>$10)} &    &$\mu_i$    &-1.21&-0.70&-0.24& 0.18& 0.67 \nl
                  &    &$\sigma_i$ & 0.23& 0.12& 0.14& 0.13& 0.19 \nl
\cline{4-8}
-3.0 - -1.0& 14& &   2&  11&   1&   0&   0 \nl
-1.0 - -0.5& 32& &   4&  15&  10&   3&   0 \nl
-0.5 -  0.0&130& &   7&  18&  77&  26&   2 \nl
 0.0 -  0.5& 73& &   1&  12&  26&  32&   2 \nl
 0.5 -  1.6& 21& &   2&   1&   2&  11&   5 \nl
\cline{4-8}
total& 270&$\mu_o$    &-0.23&-0.38&-0.13& 0.18& 0.53  \nl
     &    &$\sigma_o$ & 0.57& 0.50& 0.34& 0.33& 0.49 \nl
\enddata
\end{deluxetable}

\begin{deluxetable}{crcrrrrr}
\tablenum{4b}
\tablecolumns{7}
\tablewidth{0pt}
\tablecaption{Correlation statistics for the disk scale length. \label{tbl-4b}}
\tablehead{
\colhead{output} & \colhead{} & \colhead{} &
\multicolumn{4}{c}{input log$r_{e,D}$}  \\
\cline{4-7}
\colhead{log $r_{e,D}$} & \colhead{$N_m$} & \colhead{} & 
\colhead{-1.0 - -0.5} & \colhead{-0.5 -  0.0} & 
\colhead{ 0.0 -  0.5} & \colhead{ 0.5 -  1.6} \\
}
\startdata
{(S/N$>$200)}&  &$\mu_i$   &-0.58&-0.15& 0.31& 0.69 \nl
           &    &$\sigma_i$& 0.08& 0.12& 0.12& 0.18 \nl
\cline{4-7}
-1.0 - -0.5&  1& &   0&   0&   1&   0 \nl
-0.5 -  0.0& 75& &   2&  24&  49&   0 \nl
 0.0 -  0.5&215& &   0&   0& 163&  52 \nl
 0.5 -  1.6&108& &   0&   6&   3&  99 \nl
\cline{4-7}
total& 399&$\mu_o$    &-0.10&-0.02& 0.22& 0.58 \nl
     &    &$\sigma_o$ & 0.10& 0.46& 0.19& 0.19 \nl
\tableline
{(200$>$S/N$>$100)}&  &$\mu_i$   &-0.52&-0.20& 0.29& 0.66 \nl
                 &    &$\sigma_i$& 0.02& 0.13& 0.13& 0.15 \nl
\cline{4-7}
-1.0 - -0.5&  4& &   0&   4&   0&   0 \nl
-0.5 -  0.0&188& &   2&  54& 132&   0 \nl
 0.0 -  0.5&385& &   0&   4& 297&  84 \nl
 0.5 -  1.6&181& &   0&  20&  10& 151 \nl
\cline{4-7}
total& 758&$\mu_o$    &-0.00& 0.11& 0.19& 0.57 \nl
     &    &$\sigma_o$ & 0.00& 0.48& 0.20& 0.17 \nl
\tableline
{(100$>$S/N$>$30)}&       &$\mu_i$&-0.61&-0.20& 0.29& 0.68 \nl
                 &     &$\sigma_i$& 0.10& 0.13& 0.14& 0.16 \nl
\cline{4-7}
-1.0 - -0.5& 10& &   0&  10&   0&   0 \nl
-0.5 -  0.0&279& &  21&  83& 175&   0 \nl
 0.0 -  0.5&496& &   1&  15& 351& 129 \nl
 0.5 -  1.6&289& &   0&  51&  27& 211 \nl
\cline{4-7}
total&1074&$\mu_o$    &-0.05& 0.20& 0.21& 0.58 \nl
     &    &$\sigma_o$ & 0.10& 0.54& 0.22& 0.18 \nl
\tableline
{(30$>$S/N$>$10)}&        &$\mu_i$&-0.63&-0.24& 0.16& 0.61 \nl
                 &     &$\sigma_i$& 0.09& 0.13& 0.12& 0.07 \nl
\cline{4-7}
-1.0 - -0.5&  9& &   0&   9&   0&   0 \nl
-0.5 -  0.0&117& &  29&  66&  21&   1 \nl
 0.0 -  0.5& 21& &   0&   7&  13&   1 \nl
 0.5 -  1.6& 72& &   0&  63&   9&   0 \nl
\cline{4-7}
total& 219&$\mu_o$    &-0.04& 0.27& 0.23& 0.20 \nl
  &    &$\sigma_o$ & 0.08& 0.58& 0.40& 0.20 \nl
\enddata
\end{deluxetable}

\clearpage

\begin{figure}[h]
\epsfxsize=6.5truein
\epsfbox{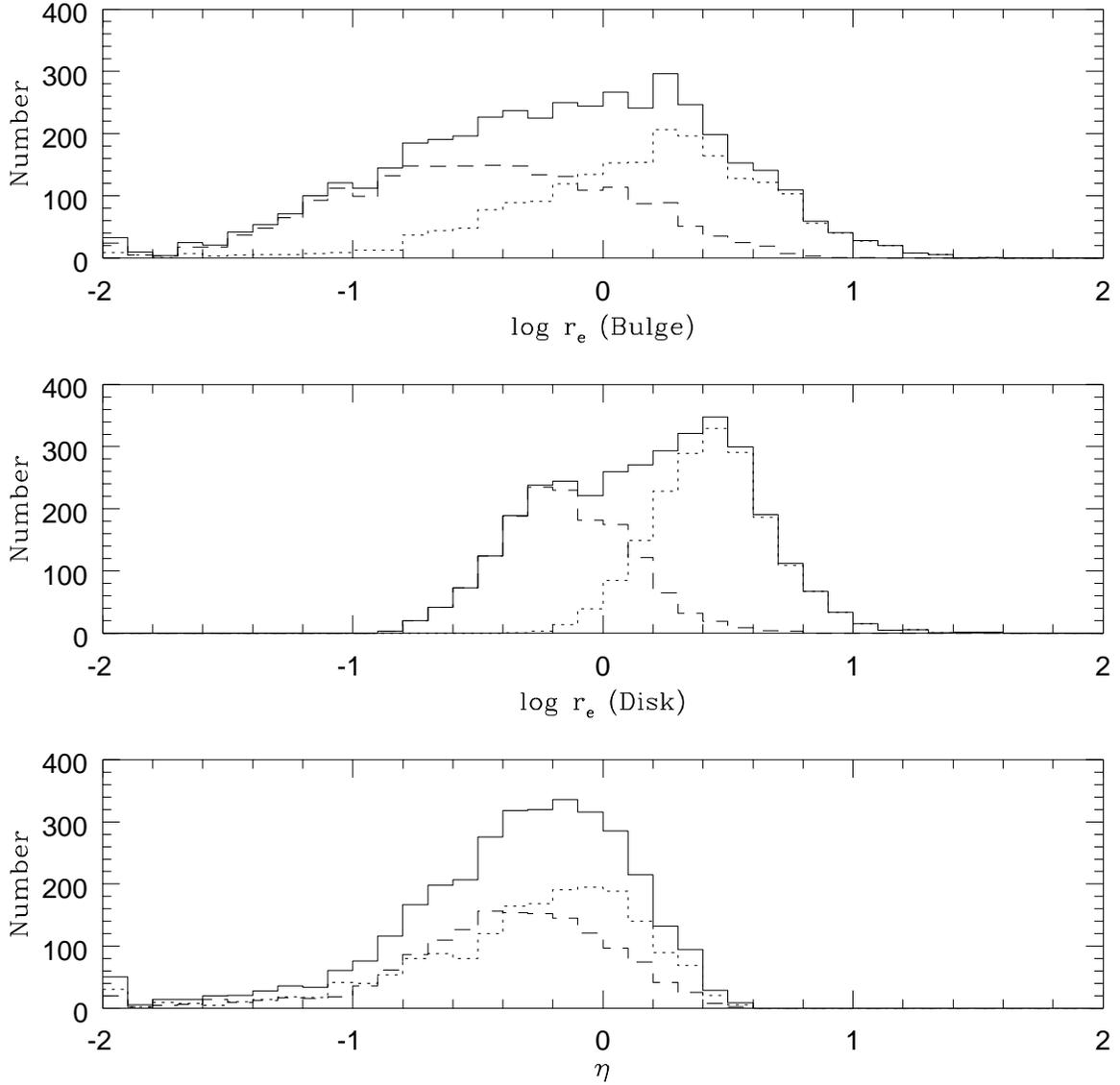}
\caption{
Histograms of $\log r_{e,B}$ (top), $\log r_{e,D}$
(middle), and $\eta=\log\left(r_{e,B}/r_{e,D}\right)$ of our sample
galaxies.  Units are in arcsec.  The solid line shows the total 4309
galaxies, while the dotted line and the broken line are for 2110
galaxies in the bright sample and for 2199 galaxies in the faint sample,
respectively.
}
\label{fig:fig1}
\end{figure}

\clearpage
\begin{figure}[h]
\epsfxsize=6.5truein
\epsfbox{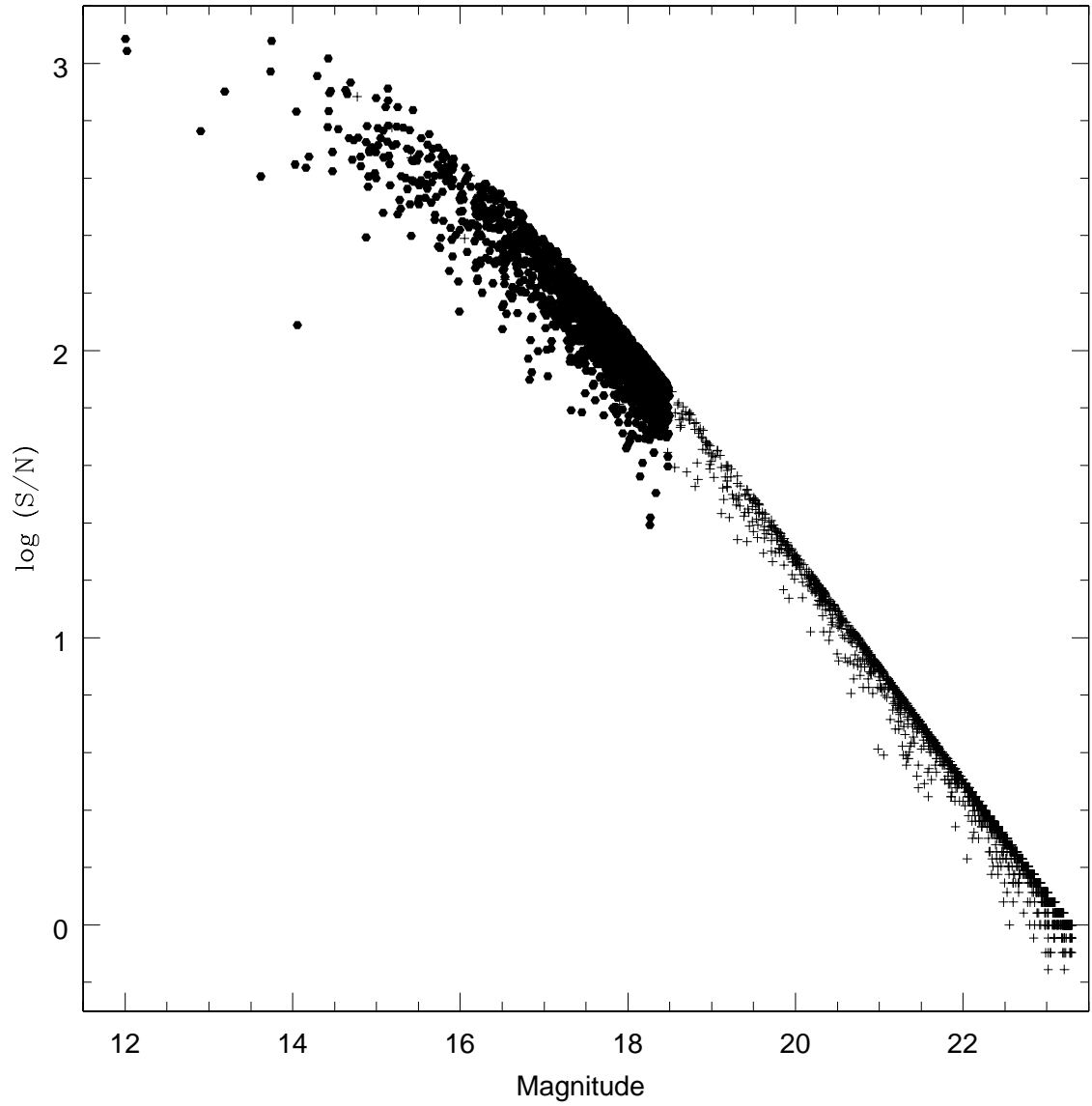}
\caption{
Signal-to-noise ratio of the 4309 simulated galaxies as a 
function of apparent total magnitude taken from the input catalog.
Filled circles are galaxies in the bright sample and pluses are 
those in the faint sample.
}
\label{fig:fig2}
\end{figure}

\clearpage
\begin{figure}[h]
\epsfxsize=6.5truein
\epsfbox{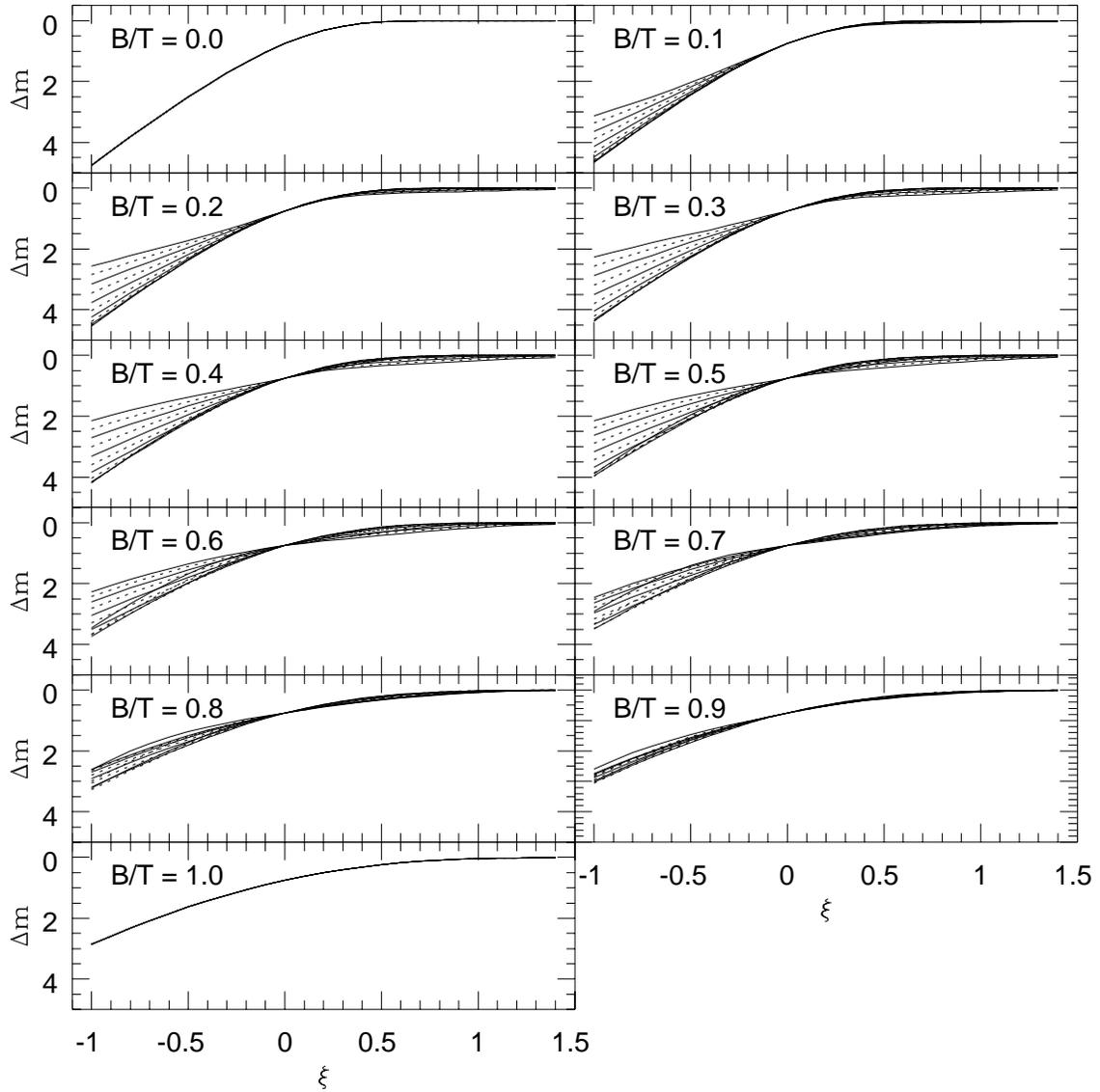}
\caption{
Template growth curves before convolution with the
seeing profile. 
}
\label{fig:fig3}
\end{figure}

\clearpage
\begin{figure}[h]
\epsfxsize=6.5truein
\epsfbox{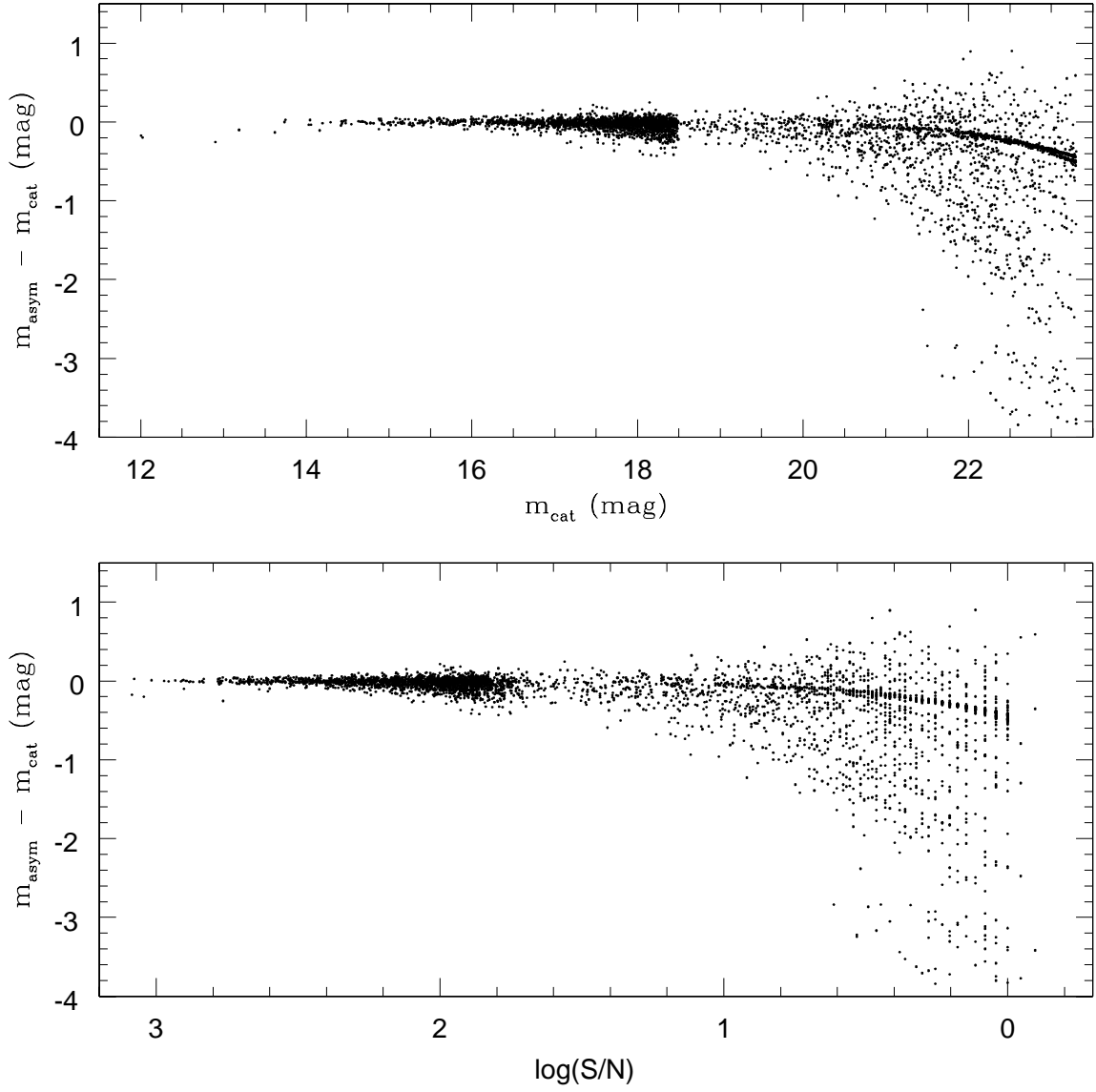}
\caption{Magnitude difference, $\Delta{m}=m_{asym}-m_{cat}$,
as a function of $m_{cat}$ (a) and of S/N (b). 
}
\label{fig:fig4}
\end{figure}

\clearpage
\begin{figure}[h]
\epsfxsize=6.5truein
\epsfbox{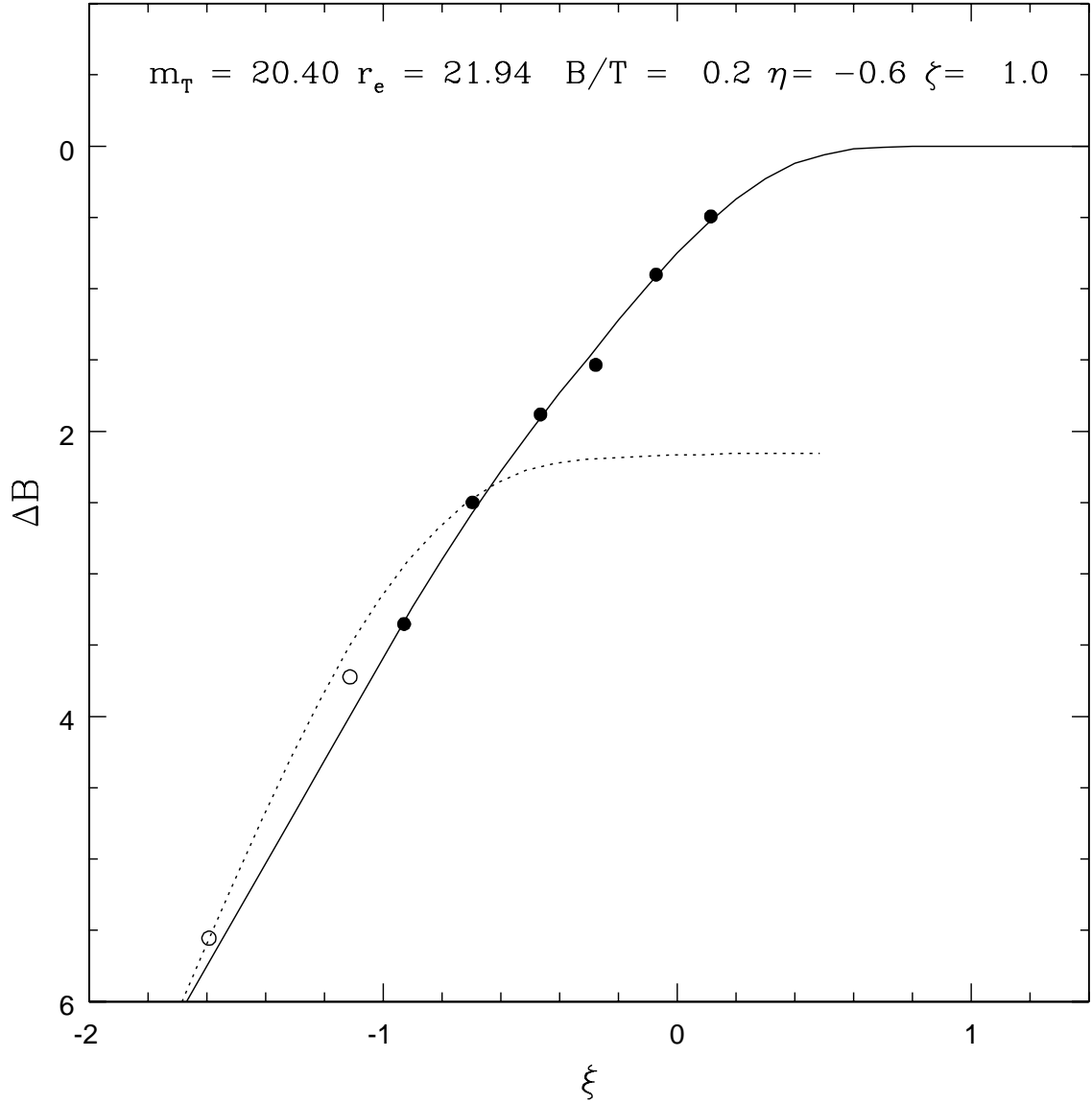}
\caption{
An example of wrong fit for a low-S/N galaxy.
The dotted curve is the growth curve of the input noiseless image. 
Open and filled circles are measurements obtained
from the simulated noisy image. The solid curve is the best fit template
(Two open circles are not used in the fit).
Outer measurements are seen to be largely affected by
noise (mostly due to incorrect background subtraction)
}
\label{fig:fig5}
\end{figure}

\clearpage
\begin{figure}[h]
\epsfxsize=2.5truein
\centerline{\epsfbox{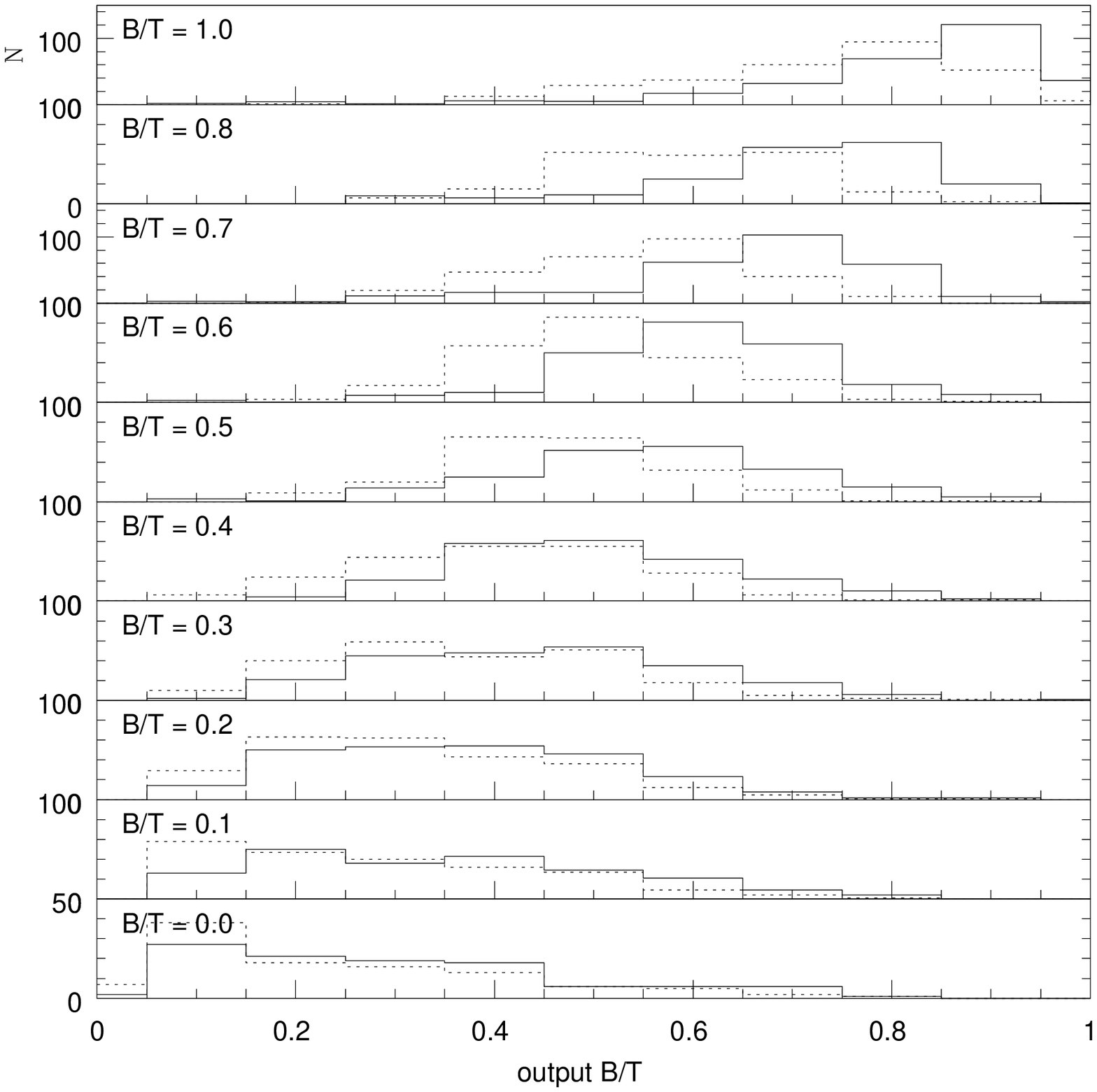}}
\epsfxsize=2.5truein
\centerline{\epsfbox{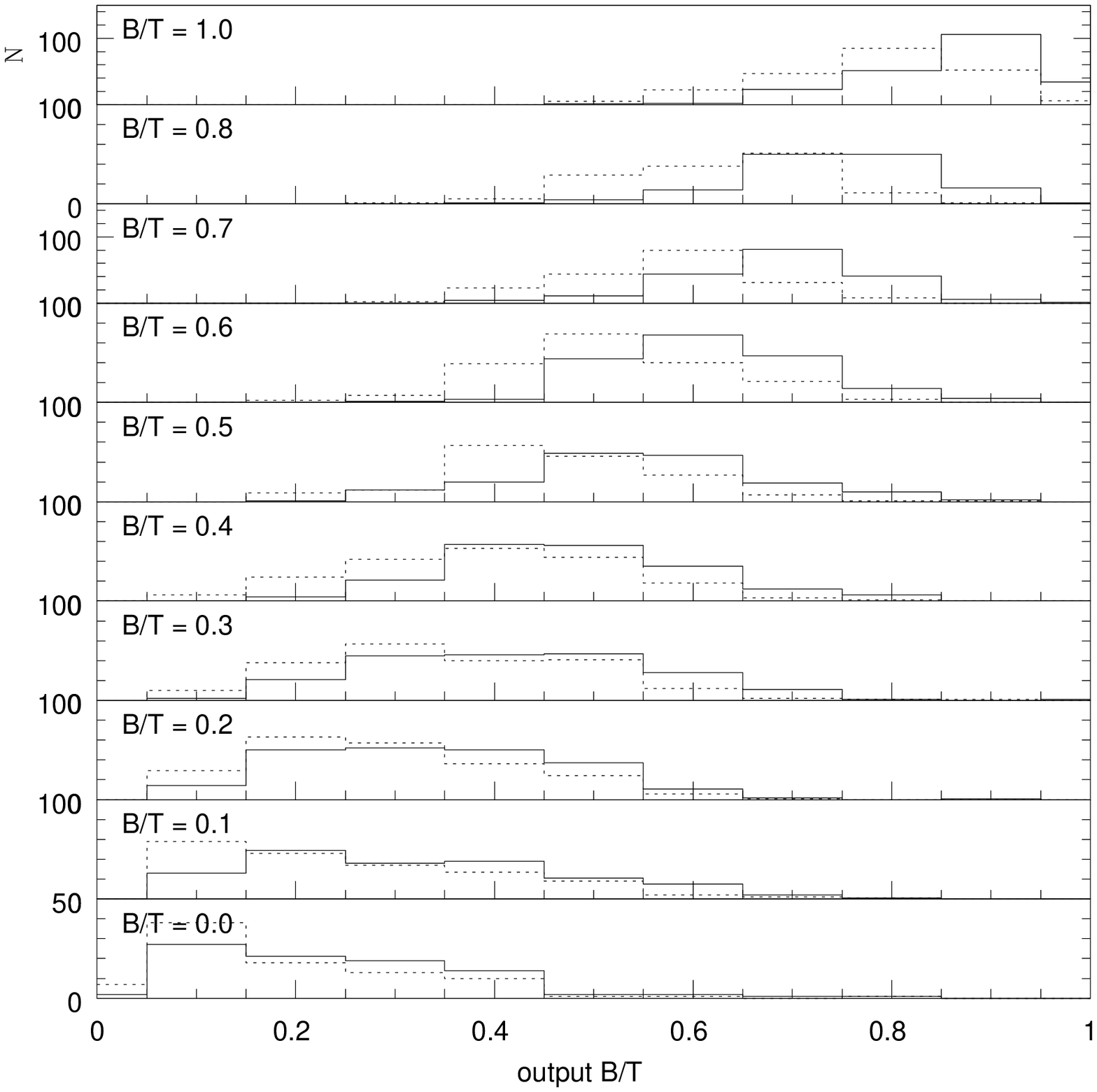}}
\epsfxsize=2.5truein
\centerline{\epsfbox{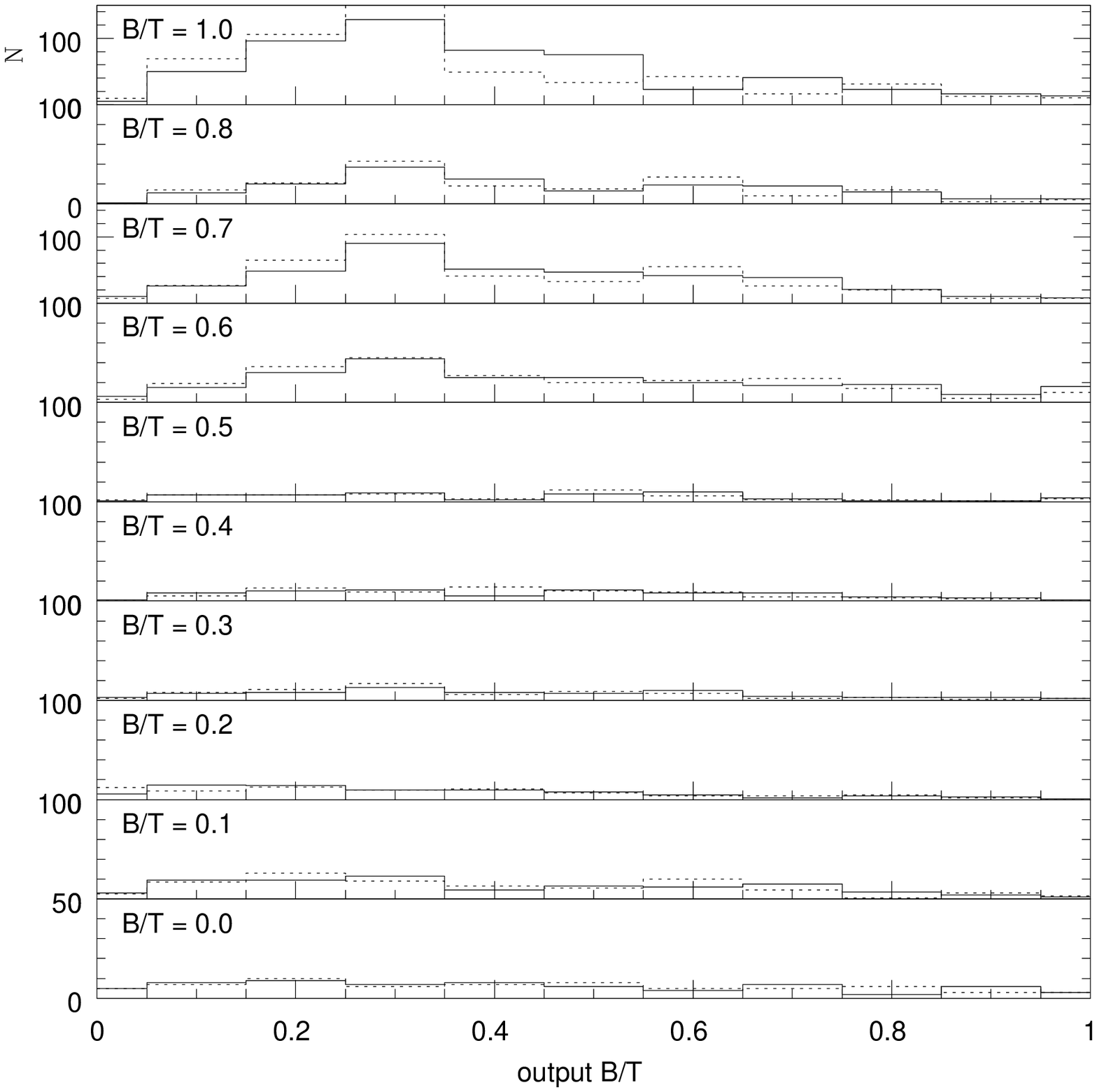}}
\caption{
The correlation between the input and output
$B/T$ ratios for all high-S/N galaxies (a), galaxies with 
S/N$>30$ and $[(B/T>0.5$ and $r_{e,B}>1)$ or
$(B/T<0.5$ and $(b/a)_D>0.2)]$(b),
and low-S/N galaxies (c).
The input (catalog) $B/T$ values are sorted into 10 bins of
width $\Delta(B/T)=0.1$, and for each of the 10 bins, the histogram
of the output $B/T$ values are shown in the separate panels.
Solid lines show the result with the true noiseless PSF
while broken lines are for artificially changed PSF
(see the text and Figure 9).
}
\label{fig:fig6}
\end{figure}

\clearpage
\begin{figure}[h]
\epsfxsize=6.5truein
\epsfbox{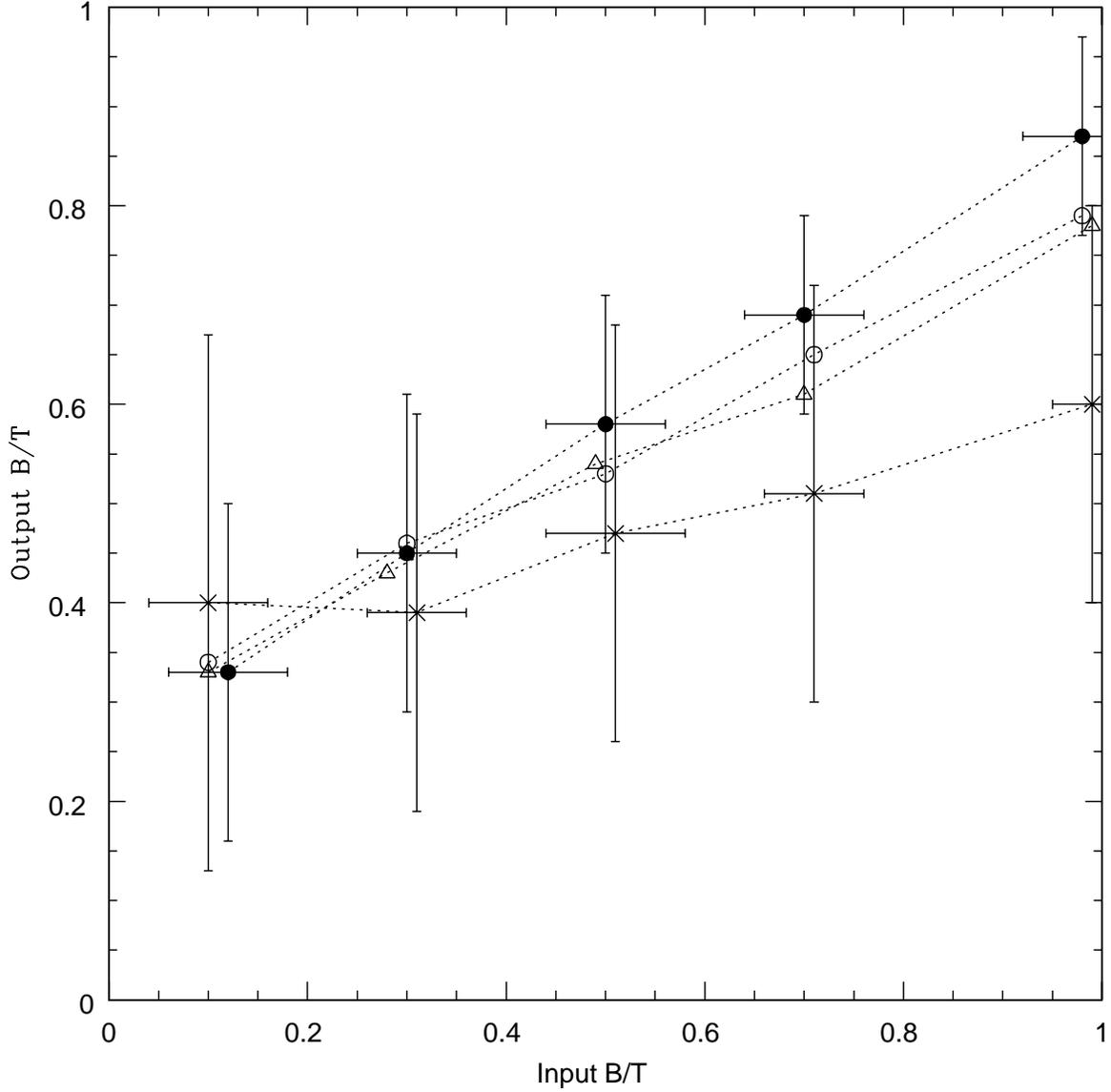}
\caption{
The mean of measured $B/T$ as a function of the 
mean of input $B/T$ for the four S/N classes. The error bars, which are 
shown only for the highest and the lowest S/N classes for clarity, 
represent the standard deviation. These values are taken from Table 3.
Filled circles, open circles, open triangles, and crosses are for
S/N$>$200, 200$>$S/N$>$100, 100$>$S/N$>$30, and 30$>$S/N$>$10,
respectively.}
\label{fig:fig7}
\end{figure}

\clearpage
\begin{figure}[h]
\epsfxsize=6.5truein
\epsfbox{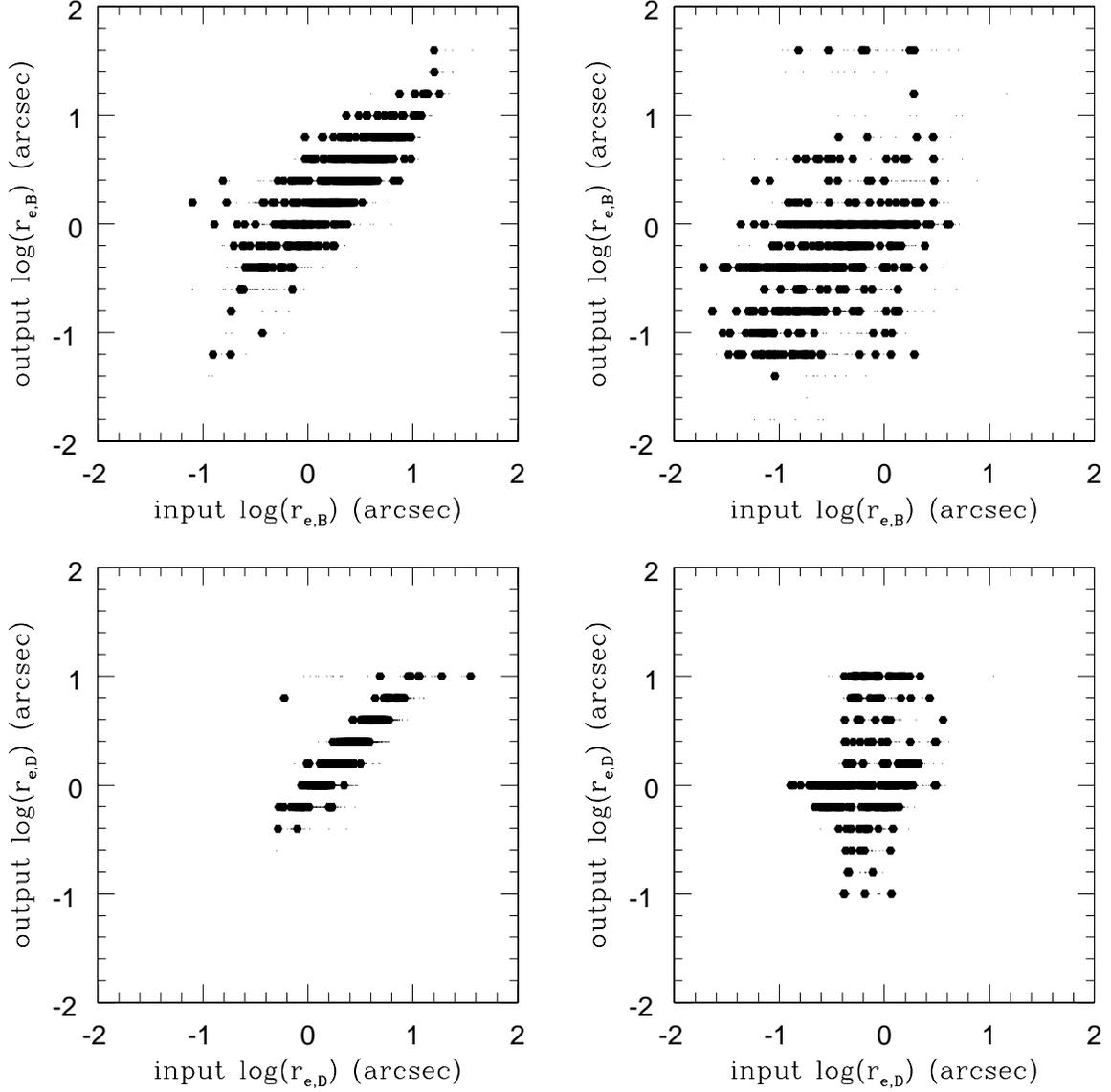}
\caption{
The correlations between the input and output values
for $r_{e,B}$ (upper panels; only for those with $B/T>$0.2)
and $r_{e,D}$ (lower panels; only for those with $B/T<$0.8).
Left panels are for high-S/N galaxies and the right panels are
for low-S/N galaxies. Filled circles are for face-on 
($(b/a)_B\geq0.8$ and $(b/a)_D\geq0.8$) galaxies and small dots are
inclined ($(b/a)_B<0.8$ or $(b/a)_D<0.8$) galaxies. 
}
\label{fig:fig8}
\end{figure}

\clearpage
\begin{figure}[h]
\epsfxsize=6.5truein
\epsfbox{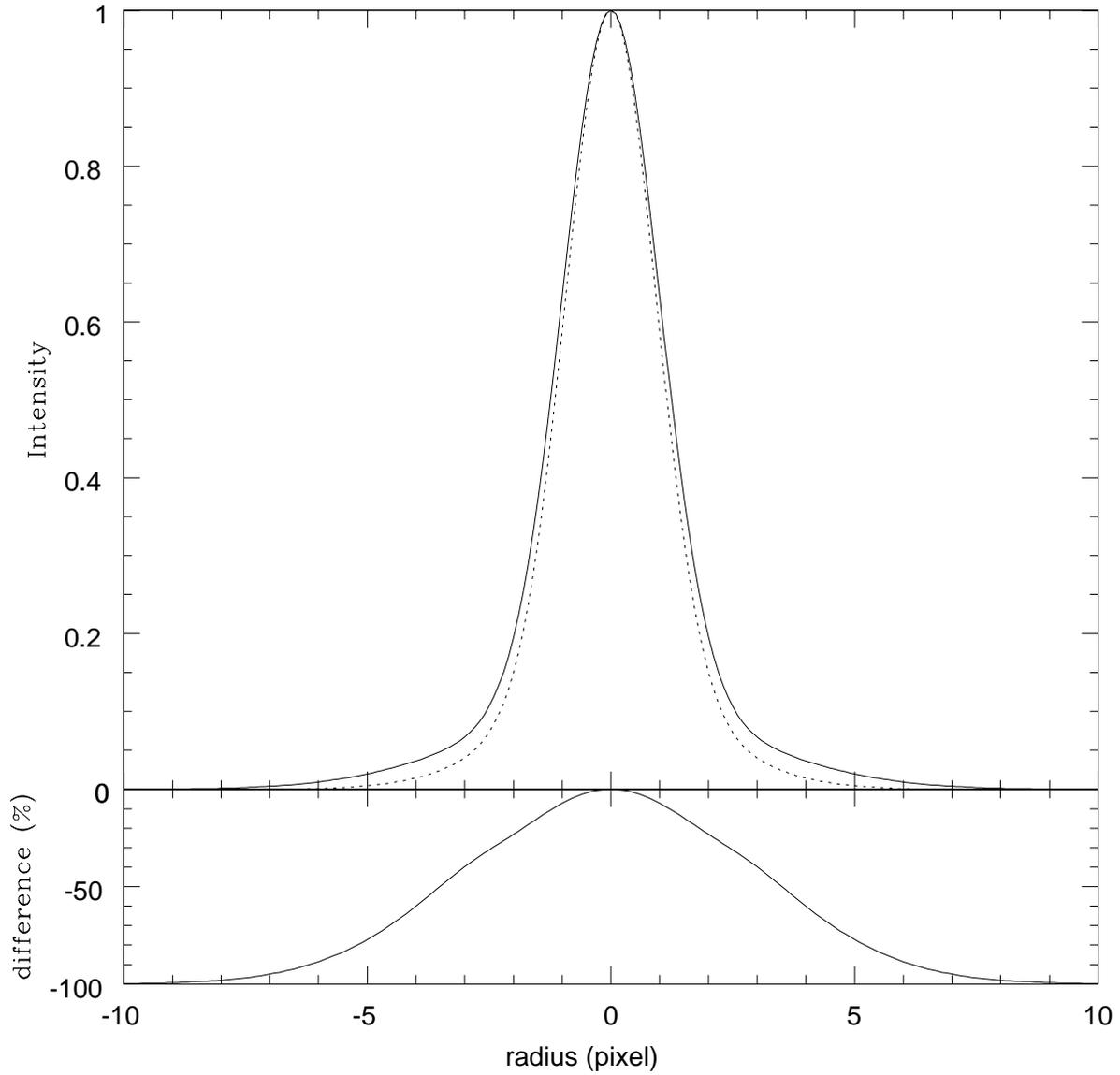}
\caption{
Upper panel shows the true noiseless PSF (solid
line) compared with the PSF reconstructed from four simulated noisy
images of a 21 mag star (broken line) (see the text). Lower panel shows
the difference, (PSF(true)-PSF(rec.))/PSF(true).
}
\label{fig:fig9}
\end{figure}

\clearpage
\begin{figure}[h]
\epsfxsize=3.5truein
\centerline{\epsfbox{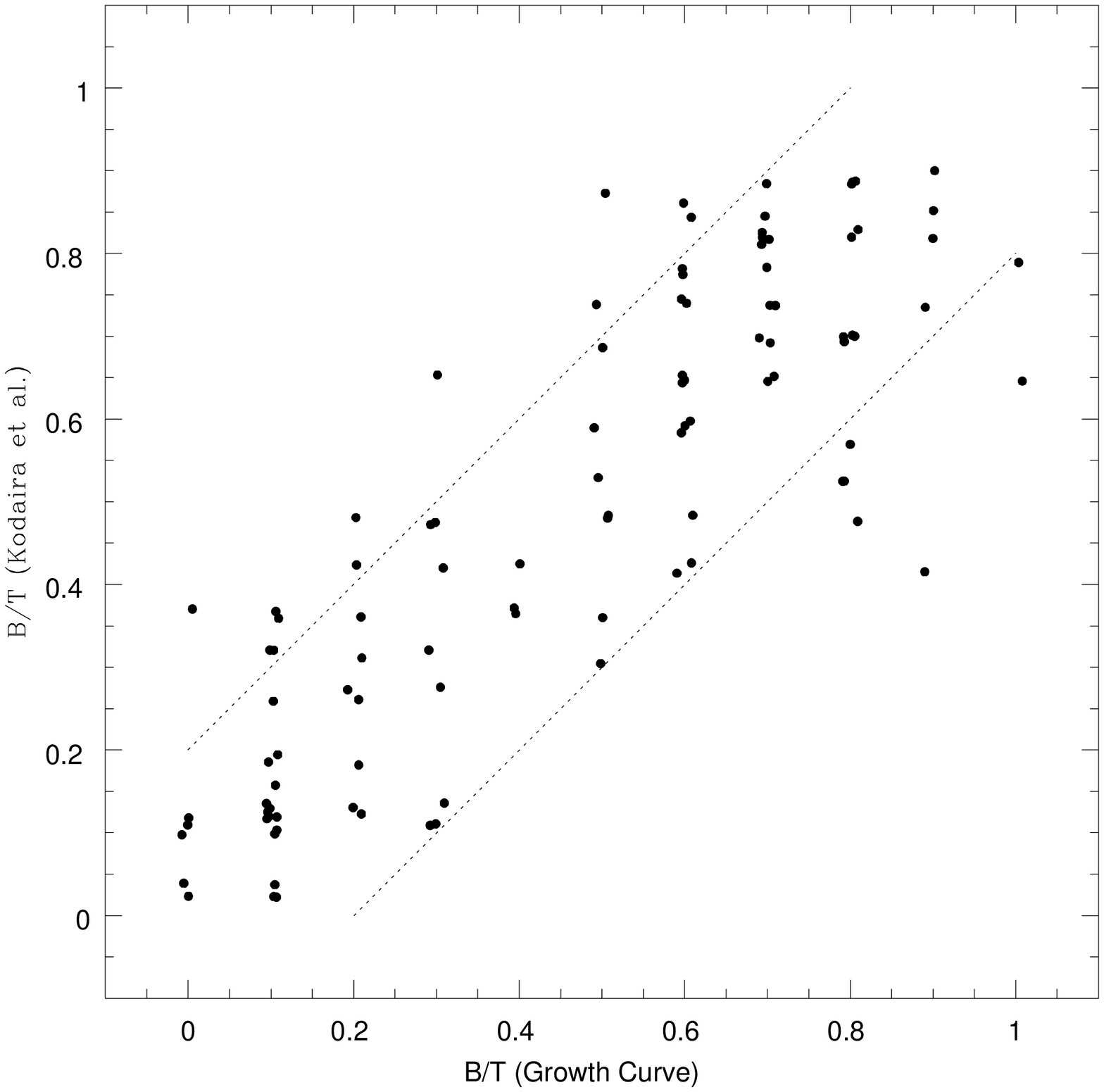}}
\caption{
Comparison of $B/T$ between Kodaira et al. (1986)
and present study. Broken line show $\Delta(B/T)=\pm$0.1.
}
\label{fig:fig10}

\epsfxsize=3.5truein
\centerline{\epsfbox{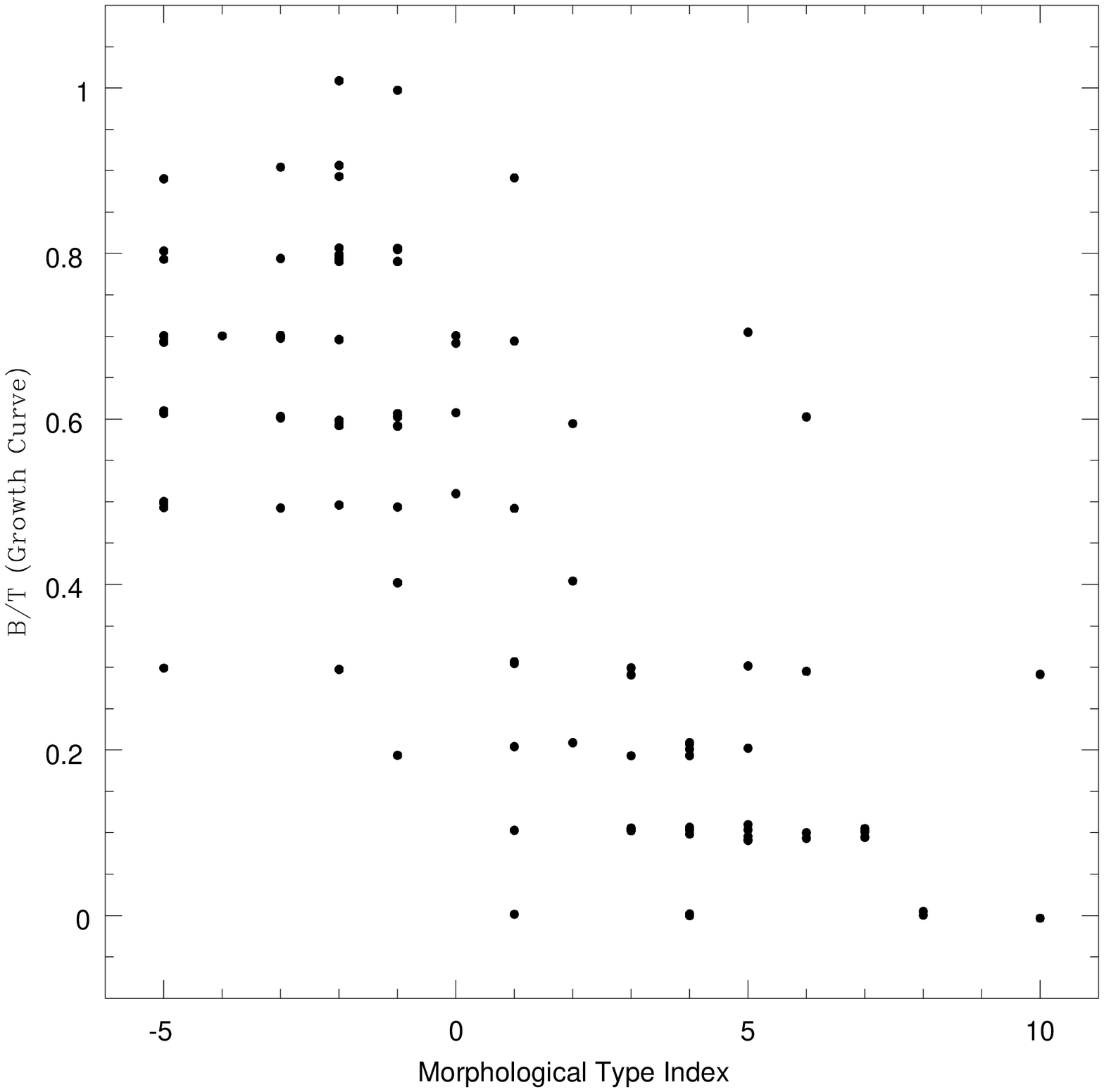}}
\caption{
$B/T$ obtained by the present method as a function of
the morphological type index
}
\label{fig:fig11}
\end{figure}

\end{document}